# Mobility-Induced Sensitivity of UAV-based Nodes to Jamming in Private 5G Airfield Networks – An Experimental Study

Pavlo Mykytyn[1], Ronald Chitauro[2], Onur Yener[1,2], Peter Langendoerfer[1,2]
[1]BTU Cottbus-Senftenberg, Cottbus, Germany
[2]IHP- Leibniz Institute for High Performance Microelectronics
mykytyn@b-tu.de

## Abstract

This work presents an experimental performance evaluation of a private 5G airfield network under controlled directional SDR jamming attacks targeting UAV-based UE nodes. Using a QualiPoc Android UE, mounted as a payload on a quadcopter UAV, we conducted a series of experiments to evaluate signal degradation, handover performance, and service stability in the presence of constant directional jamming. The conducted experiments aimed to examine the effects of varying travel speeds, altitudes, and moving patterns of a UAV-based UE to record and analyze the key physical-layer and network-layer metrics such as CQI, MCS, RSRP, SINR, BLER, Net PDSCH Throughput and RLF. The results of this work describe the link stability and signal degradation dependencies, caused by the level of mobility of the UAV-based UE nodes during autonomous and automatic operation in private 5G Airfield networks.

## 1 Introduction

Fifth generation (5G) networks, including private 5G Campus networks deployed at critical infrastructures like airfields, promise high throughput and ultra-reliable connectivity. Their deployment is becoming more dominant, as organizations try to leverage and benefit from designing and operating their own 5G solutions. The deployment of these networks is driven by the need for enhanced security, reliability, and performance in industrial and academic environments [1]. However, these networks remain susceptible to targeted radio frequency (RF) jamming attacks, enabled by low-cost Software Defined Radios (SDRs), where a malicious attacker emits interference to degrade the signal-to-noise ratio (SNR) of legitimate communication links. Recent research works such as [2], present new highly sophisticated jamming techniques that selectively target certain portions of carrier bandwidth to maximize disruption while minimizing the jammer's power and detectability. In particular, the Synchronization Signal Block (SSB) that carries the primary and secondary synchronization signals and Physical Broadcast Channel (PBCH) with critical system information are known weak points. Attacks on the SSB can block user equipment (UE) from accessing periodic synchronization, effectively causing denial-of-service on the cell [3]. On the other hand, with a powerful transmitter, full-band barrage jamming, can still be used to overwhelm the 5G downlink.

Unmanned aerial vehicles (UAVs) introduce new dimensions to this threat landscape. UAVs can serve both as tools for network monitoring and as potential jamming platforms. Prior works, such as [4], have even proposed mounting low-power jammers on small drones to hover near base stations and disrupt 5G signals in a stealth manner. At the same time, equipping drones with cellular test equipment allows researchers to probe network performance in the air and prepare for the deployment of UAVs as the 5G Nodes. However, high node mobility in 5G networks poses additional challenges, such as problematical handover procedures and shortened connection time within single cell tower range. Previous works, such as [5], have even attributed the handover failures in 5G networks to factors such as high UE mobility, intentional interference, and signal unreachability.

The integration of UAVs with private 5G networks at airfields and other areas with limited access, promises enhanced connectivity and high data throughput rates to improve user experience, security, and area management applications [6]. However, the high mobility of UAVs and their exposure to adversaries raise critical concerns about link stability, especially at the network edge. Among the most critical threats is targeted jamming, which can disrupt both initial access and ongoing sessions, potentially compromising critical services of an airfield.

The interchange between high node mobility and jamming effectiveness is thus an important open question. While studies, such as [1], [6] offer valuable insights into the theoretical aspects of 5G security and related simulations, they do not involve experimental testing or validation of such use cases in the real-world scenarios.

Our research work aims to fill this gap by experimentally testing directional jamming attacks focused selectively at the SSB versus PBCH portions of the communication spectrums under normal communication conditions and cell tower handover scenarios using a constant barrage jammer. We evaluate the signal degradation and link stability using key physical-layer and link-layer metrics. The scope of this research work, however, is limited to experimental testing and evaluation of the influence of flight speed of UAV-based UEs on the effectiveness of a barrage jammer, targeting the PBCH segments of the 5G communication spectrum.

The rest of the paper is structured as follows: In Section 2 we present our experimental setup and methodology. In Section 3 we describe the obtained results, and in Section 4 we conclude this work.

## 2 Experimental Setup

The experiments were conducted on a private 5G Campus network deployed at a small fully operational airfield. The

gNodeB of the base station and the network were configured for wide-area coverage of the airfield with two active cell towers operating in a sub-6 GHz 5G band.

## 2.1 UAV-based QualiPoc UE

A quadcopter UAV was fitted with a Rohde&Schwarz QualiPoc Android network quality testing device representing the UE to measure the 5G performance and collect physical-layer and link-layer key signal quality metrics in real time during an automatic mission flight as presented in the Figure 1.

To emulate realistic usage patterns and generate network load, two distinct traffic scenarios were developed during the UAV measurement campaigns. In the first scenario, a large downlink-dominant use case was simulated by initiating a download of a 10 GB CentOS image file, resulting in sustained high-volume downlink traffic. In the second scenario, a balanced uplink and downlink traffic profile was established by maintaining a live video conference call using GoogleMeet. These two scenarios enabled the evaluation of jamming and mobility effects under both asymmetric and symmetric traffic conditions.

The UAV was programmed to fly multiple automatic missions with a constant altitude of 11 m and varying speed of 3 m/s, 6 m/s, and 12 m/s, depending on the mission. In each conducted experiment, the UAV followed a constant preprogrammed mission flight path, covering an area of ~2 $km^2$ as depicted in the Figure 2.

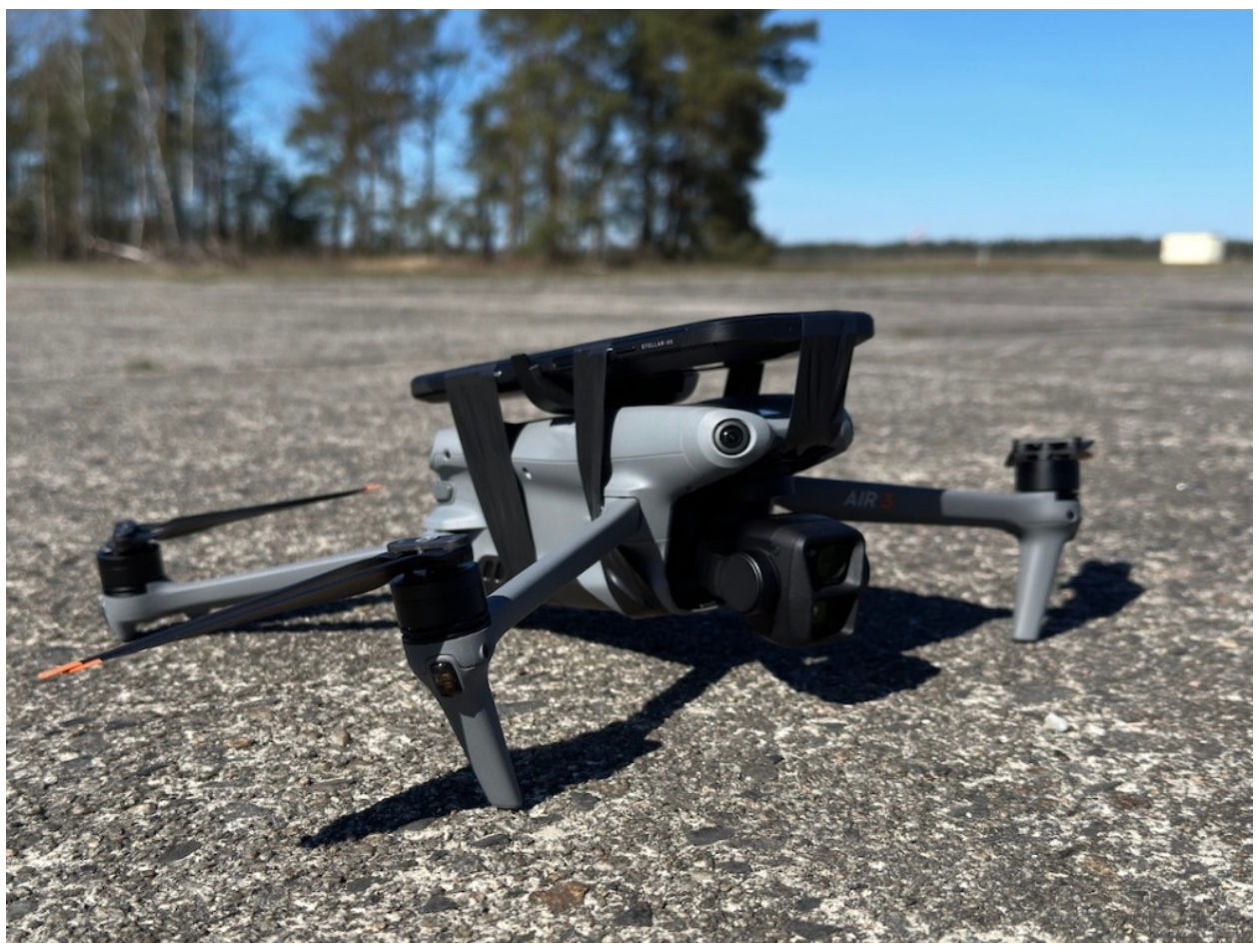

**Figure 1** Quadcopter UAV fitted with a Rohde&Schwarz QualiPoc Android network quality testing device.

## 2.2 Jamming Hardware and Software

A constant directional RF jammer was set up on the ground in the center of the area covered by the automatic flight mission as depicted in the Figure 3. The hardware utilized for jamming in the experiments consisted of a Linux-based Dell Latitude E5470 laptop running Ubuntu 22.04 and a programmable USRP B210 SDR from Ettus Research. The USRP B210 supports frequency coverage from 70 MHz to 6 GHz, with an instantaneous bandwidth of up to 56 MHz and a maximum output power between 10 and 17 dBm. Additionally, the jammer was equipped with a high-gain directional antenna, aimed into the area of UAV operation, and an external power amplifier.

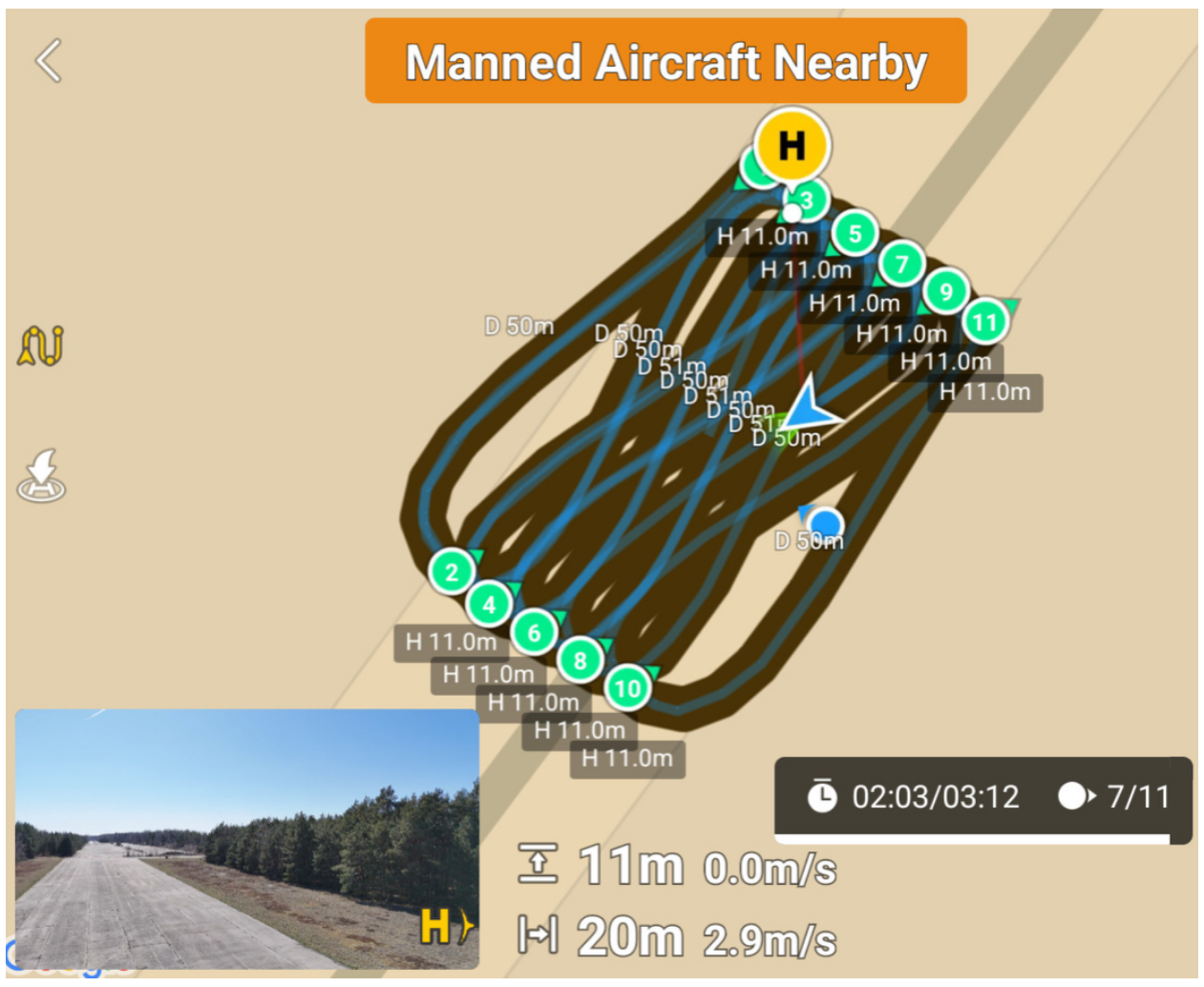

**Figure 2** Preprogrammed automatic UAV flight mission path with a constant speed and altitude.

The transmission frequency was configured to target the SSB, in a partial-band jamming mode and PBCH portions of the 5G communication channel in the barrage jamming mode. The jammer transmitted a continuous pseudo-noise signal spanning a bandwidth between 10–40 MHz and transmission power between 20–30 dBm, depending on the targeted portion of the spectrum. To ensure the jammer was working correctly, during each experiment we recorded a spectrogram of the communication channel using a Rohde&Schwarz FSH8 spectrum analyzer.

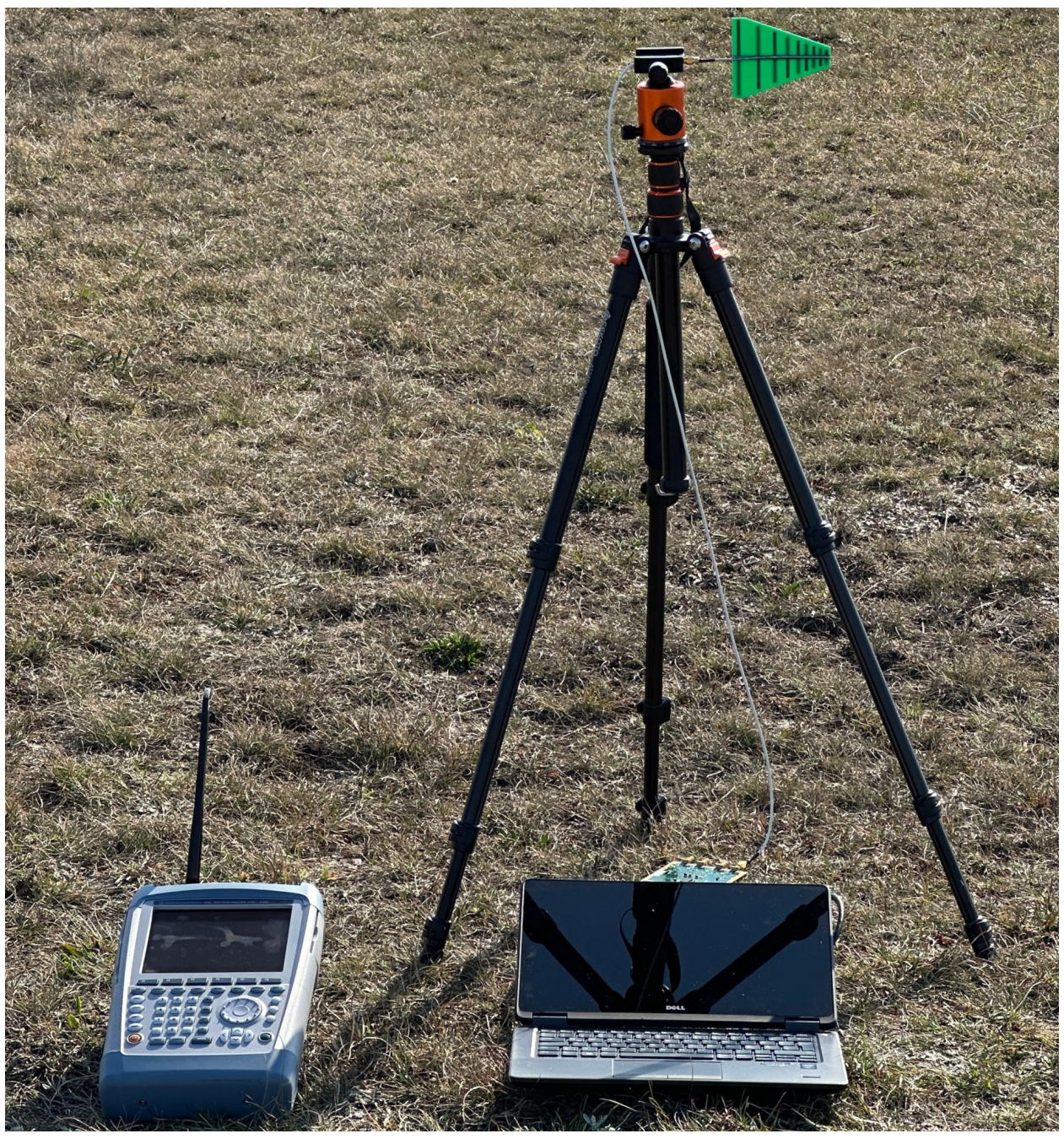

**Figure 3** SDR-based jamming hardware with a high-gain directional antenna and Rohde&Schwarz FSH8 spectrum analyzer.

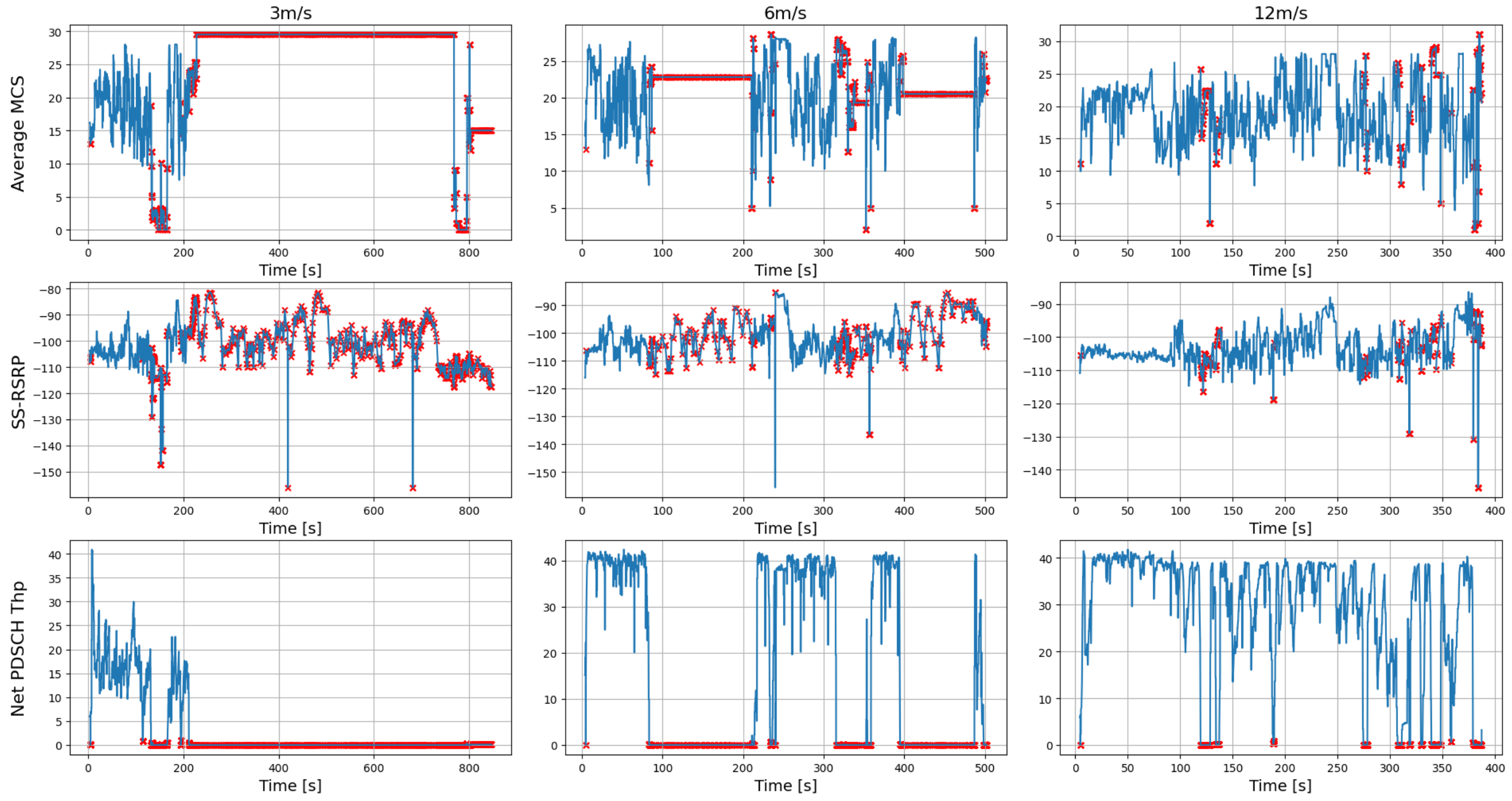

**Figure 4** Comparison of the selected 5G channel quality metrics across different UAV travel speeds collected during the data link jamming campaign

# 3 Experimental Campaigns and Results

## 3.1 SSB Jamming Campaign

The first part of our jamming experiments, focused on partial-band jamming, targeted the portion of the 5G communication channel carrying the SSB frames. This measurement experiment was conducted on the ground before conducting the UAV flight missions. The targeted SSB bandwidth was 7.2 MHz wide, with the center frequency at 3743.10 MHz. The UE receiver was positioned 100 m away from the base station cell tower, and the jammer was placed at the mid-point between the UE and the cell tower, with its antenna aimed at the UE. During this measurement campaign, we initiated a download of a 10 GB CentOS image file, resulting in continuous downlink traffic. During this experiment we observed that jamming of the SSB portion of the spectrum was extremely successful. After jammer activation in each of the measurement sessions, the downlink traffic was completely disrupted, and the UE was disconnected from the base station. The throughput values, therefore, went down to zero.

## 3.2 Data Link Jamming Campaign

The full band jamming targeted the PBCH part of the 5G communication channel responsible for uplink and downlink communication. The SSB band was not affected during this jamming campaign. The targeted PBCH bandwidth was 60.8 MHz wide, with the center frequency of 3766.05 MHz. The jammer was configured to target a 20 MHz portion of the bandwidth and rapidly and repeatedly sweep through the 60.8 MHz frequency band to accomplish the jamming attack. The measurement sessions during this campaign were structured into 3 jamming experiments. The UAV was programmed to fly three separate identical missions at a constant altitude of 11 m and speed varying between 3 m/s, 6 m/s, and 12 m/s. To balance the uplink and downlink traffic during this experimental campaign, a live video conference call was initiated on the UE using GoogleMeet and maintained throughout all conducted experiments. The UE mounted to the UAV was programmed to collect relevant signal quality metrics and measure signal degradation under active data link jamming.

## 3.3 Signal Quality Metrics and Experimental Results

During the data link jamming campaign, we collected the Modulation and Coding Scheme Index (MCS), Reference Signal Received Power (RSRP), Signal-to-Interference-plus-Noise Ratio (SINR), Wideband Channel Quality Indicator (CQI WB), Block Error Rate (BLER), Retransmission count, and Net PDSCH Throughput values.

MCS is a unitless index representing the modulation type and coding rate applied to the physical downlink and uplink transmissions. Higher MCS reflects good channel conditions and lower MCS indicates fallback to low-throughput schemes under poor signal conditions. Monitoring MCS over time helps to evaluate the dynamic adaptability of the 5G link and its throughput efficiency. RSRP is measured in dBm and reflects the average received power of the synchronization signal, calculated at the UE. It indicates raw signal strength and is crucial for cell selection, handover, and beam management. Net PDSCH throughput reflects the actual UE data throughput received on the downlink, after accounting for retransmissions and protocol overhead and is a primary indicator of service quality.

Figure 4 represents the comparison between average MCS, RSRP, and PDSCH Throughput across different UAV speeds plotted over time. Each column corresponds to a fixed mission speed, while each row displays one of the three metrics. The red markers indicate the moments when downlink throughput fell below 1 Mbps, highlighting significant degradation of the communication link quality. At the speeds of 3 m/s and 6 m/s, the average MCS was higher than that of a 12 m/s, yet both 3 m/s and 6 m/s experiments experienced complete outages of the received signal caused by the jamming effects, while at 12 m/s, though with lower MCS values, the signal remained uninterrupted. While analyzing the RSRP across three different speeds, the values between experiments remained comparable, and did not show significant discrepancies, variating between - 90 and -110 dBm. This confirmed our expectations, because during the data link jamming campaign, only the PDSCH bandwidth was affected by jamming, leaving the SSB portion of the spectrum interference free.

While analyzing the Net Throughput across different speeds, the service quality variations between the speeds became even more clear. At 3 m/s and 6 m/s, the average recorded throughput was higher, however, the signal cut out in multiple places, with throughput dropping to absolute zero. Although the average throughput at 12 m/s was slightly lower, the connection always stayed active, avoiding complete service disruptions and thus, reducing the overall effect of the jamming signal on the link quality.

# 4 Conclusion

The results suggest that UAV mobility has an observable impact on the effectiveness of jamming attacks and 5G communication performance. Faster moving speed reduced the time the UAV spent inside high-interference zones, helping the UE escape jamming hotspots before link conditions could collapse.

However, it also had a negative effect on the 5G link stability and overall recorded throughput. Constant targeted jamming was very effective in all of the experiments, however, its impact on UE was most severe at lower traveling speeds. Higher UAV speeds, despite the lower average recorded MCS and throughout values, increased the resilience of the 5G link to targeted, directional and stationary jamming attacks by reducing the exposure time of UE to peak interference.